\documentclass[aps,twocolumn,
groupedaddress,nofootinbib,nobalancelastpage,nobibnotes]{revtex4}
\pdfoutput=1
\usepackage{amsmath,amsfonts,amssymb,mathrsfs,graphicx}
\usepackage[squaren]{SIunits}

\newcommand{\ie}{{\it i.e.}}

\newcommand{\eg}{{\it e.g.}}

\newcommand{\eq}{Eq.}

\newcommand{\fig}{Fig.}

\newcommand{\Ref}{Ref.}
\newcommand{\Refs}{Refs.}

\newcommand{\Tab}{Tab.}

\newcommand{\equ}[1]{\eq~(\ref{equ:#1})}
\newcommand{\figu}[1]{\fig~\ref{fig:#1}}

\newcommand{\bi}{\begin{itemize}}
\newcommand{\ei}{\end{itemize}}

\begin{document}

\preprint{DESY 14-136}

\title{Describing the Observed Cosmic Neutrinos by Interactions of Nuclei with Matter}

\author{Walter Winter}
\affiliation{DESY, Platanenallee 6, 15738 Zeuthen, Germany}

\date{\today}

\begin{abstract}
IceCube have observed neutrinos which are presumably of extra-galactic origin. Since specific sources have not yet been identified, we discuss what could be learned from the conceptual point of view. We use a simple model for neutrino production from the interactions between nuclei and matter, and we focus on the description of the spectral shape and flavor composition observed by IceCube. Our main parameters are spectral index, maximal energy, magnetic field, and composition of the accelerated nuclei. We show that a cutoff at PeV energies can be achieved by soft enough spectra, a cutoff of the primary energy, or strong enough magnetic fields. These options, however, are difficult to reconcile with the hypothesis that  these neutrinos originate from the same sources as the ultra-high energy cosmic rays. We demonstrate that heavier nuclei accelerated in the sources may be a possible way out if the maximal  energy scales appropriately with the mass number of the nuclei. In this scenario, neutrino observations can actually be used to test the UHECR acceleration mechanism. We also emphasize the need for a volume upgrade of the IceCube detector for future precision physics, for which the flavor information becomes a statistical meaningful model discriminator as qualitatively new ingredient.
\end{abstract}

\maketitle

\section{Introduction}

One of the major breakthroughs in astronomy has been the discovery of high energy neutrinos  by the IceCube detector at the South Pole. The story started with two PeV neutrino events announced in 2012~\cite{Aartsen:2013bka} as a consequence of changing the search strategy, over 28~events with deposited energies greater than 30~TeV announced in 2013~\cite{Aartsen:2013jdh}, to 37~events in the current three year data analysis~\cite{Aartsen:2014gkd}. While about 15 of these 37 events are expected from the backgrounds of atmospheric muons and neutrinos, the excess over the background currently constitutes more than $5\sigma$. Clearly, high-energy neutrino astronomy is emerging as a new discipline with IceCube at the forefront, which is expected to collect 100-200 events (within the current analysis scheme) during its lifetime. While this is significant statistics, it may not be sufficient for resolving individual sources~\cite{Ahlers:2014ioa} and for precision studies of spectrum and flavor composition. Therefore, the next generation experiments are being discussed, such as a volume upgrade of the IceCube detector (IceCube high-energy extension HEX) and the KM3NeT experiment in the Mediterranean~\cite{Katz:2006wv}. The results of the IceCube experiment during this and the coming years will have to serve as input for the optimization of these future options.

On the theoretical side, there have been many speculations where these neutrinos would be coming from, see \eg\ \Ref~\cite{Anchordoqui:2013dnh} for a recent review. It is probably fair to say that there is no general answer on that question yet. Instead, the current state-of-the-art can be recast in a number of conceptual questions: 
\begin{itemize}
 \item 
  What is the role of atmospheric neutrinos, especially prompt neutrinos?
 \item 
  Are there any directional or time-wise clusters, or is the flux isotropic? Are there any correlations with known objects or events? 
 \item 
  Are some of the events of Galactic origin?
 \item 
  Why are there no events above a few PeV?
 \item
  Can the neutrinos stem from the sources of the ultra-high energy cosmic rays?
  \item
  Are the observed neutrinos coming from one source class or more? Which ones?
 \item
  Is the flavor composition what is expected, or are there deviations indicative for either physics beyond the Standard Model or non-conventional compositions at the source?
 \item 
  Is there a particle physics origin of these neutrinos, such as dark matter?
\end{itemize}

In this study, we focus on the conceptual interpretation of the observed events in terms of the spectral shape of the observed flux. That is, we assume that the neutrinos come from one source population with similar properties which is of extra-galactic origin, as there is not yet any evidence for directional clusters. We furthermore  assume that the source population is cosmologically distributed such that it roughly follows the star formation rate. We postulate that the neutrinos are produced from interactions between nuclei and matter. This is, in a way, the simplest possible class of models, as the neutrino spectrum directly follows the non-thermal spectrum of the accelerated nuclei; for interactions with radiation, the obtained spectral shape of the neutrinos depends on both the spectra of the interacting nuclei and the target photons, see \eg\ \Ref~\cite{Winter:2013cla} for target photons produced by synchrotron radiation of co-accelerated electrons. However, we do take into account magnetic field effects on the secondary muons and pions, which can significantly alter the neutrino spectra and flavor composition, see \eg\ \Refs~\cite{Kachelriess:2007tr,Lipari:2007su,Hummer:2010ai} especially in the context of gamma-ray bursts~\cite{Kashti:2005qa,Murase:2005hy,Baerwald:2011ee} and microquasars~\cite{Reynoso:2008gs,Baerwald:2012yd}; see \Ref~\cite{Winter:2012xq} for a review.
While most of these studies discuss interactions between protons and photons, the secondaries produced by interactions between nuclei and matter will be affected by magnetic fields as well, see \eg\ \Ref~\cite{Reynoso:2008gs}. In the context of a possible cutoff at PeV energies in recent IceCube observations, magnetic field effects on the secondaries may be a way to decouple the maximal proton from the maximal neutrino energy, see \Ref~\cite{Winter:2013cla} for a more detailed discussion. Finally, there seems to be condensing evidence for a heavier composition of the UHECRs~\cite{Abraham:2010yv}. We therefore take into account the composition of the nuclei in the sources. We especially discuss if the cutoff at PeV neutrino energies can be consistent with the UHECR paradigm if heavier nuclei are accelerated to higher energies within the sources. In fact, we will demonstrate that one can learn something about the acceleration mechanism in that scenario. Note that the interpretation of the obtained neutrino flux normalization in terms of source luminosity and column depth will be discussed elsewhere.

\section{Model and methods}

\begin{table}[t]
\begin{center}
 \begin{tabular}{lll}
  \hline
  Parameter & Description & Unit  \\
  \hline
  $\alpha$ & Spectral index of primary nuclei & none \\
  $E_{\mathrm{max}}$ & Maximal energy   & GeV \\
  $B$ & Magnetic field & Gauss (G) \\
  $A$ & Mass number &  none \\
  \hline 
 \end{tabular}
\end{center}
 \caption{\label{tab:params} Main parameters of the model.}
\end{table}

The interaction model used in this work is based on the Kelner et al.~\cite{Kelner:2006tc} parameterization for proton-proton interactions, where we take into account the charged pion production explicitly to allow for secondary cooling. For the extension to heavier nuclei $Ap$ interactions, see \Ref~\cite{Joshi:2013aua}. 
The secondary production $Q_\pi$ [$\mathrm{cm^{-3} \, s^{-1} \, GeV^{-1}}$] is given from the non-thermal nucleon density in the source $N_A$ [$\mathrm{cm^{-3} \, GeV^{-1}}$] and the target nucleon density $n_p$ [$\mathrm{cm^{-3}}$] by
\begin{eqnarray}
Q_{\pi}(E_{\nu}) & = &  c  \, n_p \, \int\limits_{0}^{1} \sigma_{Ap}\left(\frac{E_\nu}{x_A} \right) \, N_A \left( \frac{E_\nu}{x_A} \right) \nonumber \\
& & \qquad \times A \, f\left(A x_A,\frac{E_\nu}{A x_A}\right) \, \frac{dx_A}{x_A} \, ,  \label{equ:int}
\end{eqnarray}
where $x_A=x/A$ is the fraction of the nucleus' energy going into the neutrino and $f$ are the scaling functions from \Ref~\cite{Kelner:2006tc} (SIBYLL-based versions) and  $\sigma_{Ap}=A^{3/4} \times \sigma_{pp}$~\cite{Anchordoqui:2006pe}. If the target material is heavier than hydrogen, one may superimpose the nuclei $n_p \simeq A \times n_A$. There are, however,  corrections to that, but, as we do not discuss the normalization in this study, that does not affect our results. Note that  \equ{int} can be re-written in terms of the  column density $L \, n_p$, where the interpretation of $L$ in terms of the size of the interaction region depends on the scenario. For example, injecting nuclei into the interaction region with the rate $Q_A$, one can estimate that $N_A \simeq Q_A \, t_{\mathrm{esc}}$ in the absence of disintegration and cooling, and therefore $L \, n_p = c \, t_{\mathrm{esc}} \, n_p$ is the column density which determines the normalization of \equ{int}.
 The pion and consequent muon decays are computed in the usual way including the helicity dependence of the muon decays, see  \Ref~\cite{Hummer:2010ai}. Flavor mixing is taken into account with the best-fit values from \Ref~\cite{GonzalezGarcia:2012sz} (first octant solution).

The main parameters of the model are listed in \Tab~\ref{tab:params}. We start with $N_A(E) \propto E^{\alpha} \, \exp(-E/E_{\mathrm{max}})$ in \equ{int}, where $\alpha$ is the initial spectral index, which is expected to be $\alpha \sim 2$ from Fermi shock acceleration. The maximal energy $E_{\mathrm{max}}$ is typically obtained from equating the acceleration timescale with the dominant escape and energy loss timescales in a specific scenario. Since this derivation is highly model-dependent, we keep $E_{\mathrm{max}}$ as a model parameter. Note that $E_{\mathrm{max}}$ can also be used to simulate a spectral break in the initial spectrum, as it may come from an energy-dependent escape time frequently discussed for starburst galaxies~\cite{Loeb:2006tw}, see also \Refs~\cite{Anchordoqui:2014yva,Chang:2014hua,Tamborra:2014xia}. In addition,  in \equ{int}, $N_A$ already corresponds to the result including disintegration and other cooling and escape processes. That means that $E_{\mathrm{max}}$ could also describe a spectral break from a cooling process, energy-dependent escape, or photo-disintegration. The secondary muons and pions are assumed to undergo synchrotron losses and decay governed by the magnetic field $B$, which impacts spectral shape and flavor composition see \Refs~\cite{Hummer:2010ai,Baerwald:2011ee}. Note that there could be other cooling or escape processes affecting the secondaries, such as adiabatic cooling (see \eg\ \cite{Hummer:2011ms}) or re-acceleration~\cite{Klein:2012ug,Winter:2014tta,Reynoso:2014yoa}. These effects are, however, model-dependent and typically not the dominant ones shaping the neutrino spectra. Finally, we have the composition $A$ as parameter. In fact, \equ{int} allows to use an energy-dependent (average) composition $A(E_A)=A(E_\nu/x_A)$, which we will use below. Note that for a power law with $\alpha=-2$, the composition would not affect our results. However, both the cutoff and varying composition will change that conclusion.

For the sake of simplicity, we furthermore assume that the sources do not have large Doppler factors, and that they are cosmologically distributed following the star formation rate by Hopkins and Beacom~\cite{Hopkins:2006bw}, \ie, $E^2 \phi_\nu \propto \int_z  Q_\nu(E(1+z)) \, H(z) \, dV/dz \, (4 \pi d_L^2)^{-2} dz$  for steady sources with $H(z)$ the source density normalized to the local source density ($H(0) \equiv 1$), $\Omega_m=0.27$ and $\Omega_\Lambda=0.73$. For transients, there will be another factor $1+z$ in the denominator of the integrand; we find however that the results depend very little on the details of the cosmological source distribution.

For the fit, we follow \Ref~\cite{Winter:2013cla}, using the up-to-date three year data from \Ref~\cite{Aartsen:2014gkd}. We use eight bins: four in the {\em reconstructed} neutrino energy 30 to 200~TeV, 200 TeV to 1 PeV, 1 to 3 PeV, and 3 PeV to 100 PeV, and one bin for muon tracks and cascades within each energy slot.  For the sake of simplicity, we assume that the electromagnetic equivalent energy is roughly 25\% of the incident neutrino energy for a muon track, and 75\% for a cascade~\cite{Laha:2013lka}; cascades from neutral current interactions are assumed to be suppressed by the cross sections, the lower fiducial mass (see Fig.~7 in \Ref~\cite{Aartsen:2013jdh}), and the fact that only a fraction of the initial neutrino energy is deposited in the detector~\cite{Aartsen:2013bka}. 
 The (Poissonian) $\chi^2$ is obtained by bin-wise comparing the observed 36~events, for which energy information is available, with the prediction. The prediction with a free overall normalization is obtained from folding the three flavored neutrino fluxes for one set of model parameters with the corresponding exposures derived from the effective areas in \Ref~\cite{Aartsen:2013jdh}. Then the atmospheric backgrounds are added. The atmospheric neutrino background is derived from the IceCube observation of muon neutrinos~\cite{Abbasi:2010ie,Abbasi:2011jx}, and for the atmospheric muons we assume the same shape. The measurement of the atmospheric electron neutrino background is much more uncertain. We therefore extrapolate it from the muon neutrino background, roughly consistent with the flavor composition in \cite{Sinegovskaya:2013wgm}. We obtain 3.6 background muon tracks from neutrinos, 3.2 cascades from neutrinos, and 8.6 muon tracks from atmospheric muons, matching the publically available information on the IceCube analysis. This means that the predicted number of muon tracks is slightly higher than the observation (eight), a fact which discussed in detail in \Ref~\cite{Mena:2014sja} (see also \Ref~\cite{Chen:2013dza} for some discussion). While this unavoidable tension increases the minimal $\chi^2$ of our fit, it hardly affects the $\Delta \chi^2$, and therefore if of little relevance for the results in this study. As a final step the $\chi^2$ between predicted and observed rates are summed over all bins and  minimized over the free normalization of the astrophysical flux to obtain the best-fit for the chosen parameters.

\section{Results for proton-matter interactions}

\begin{figure*}[t!]
\begin{center}
\includegraphics[width=0.8\textwidth]{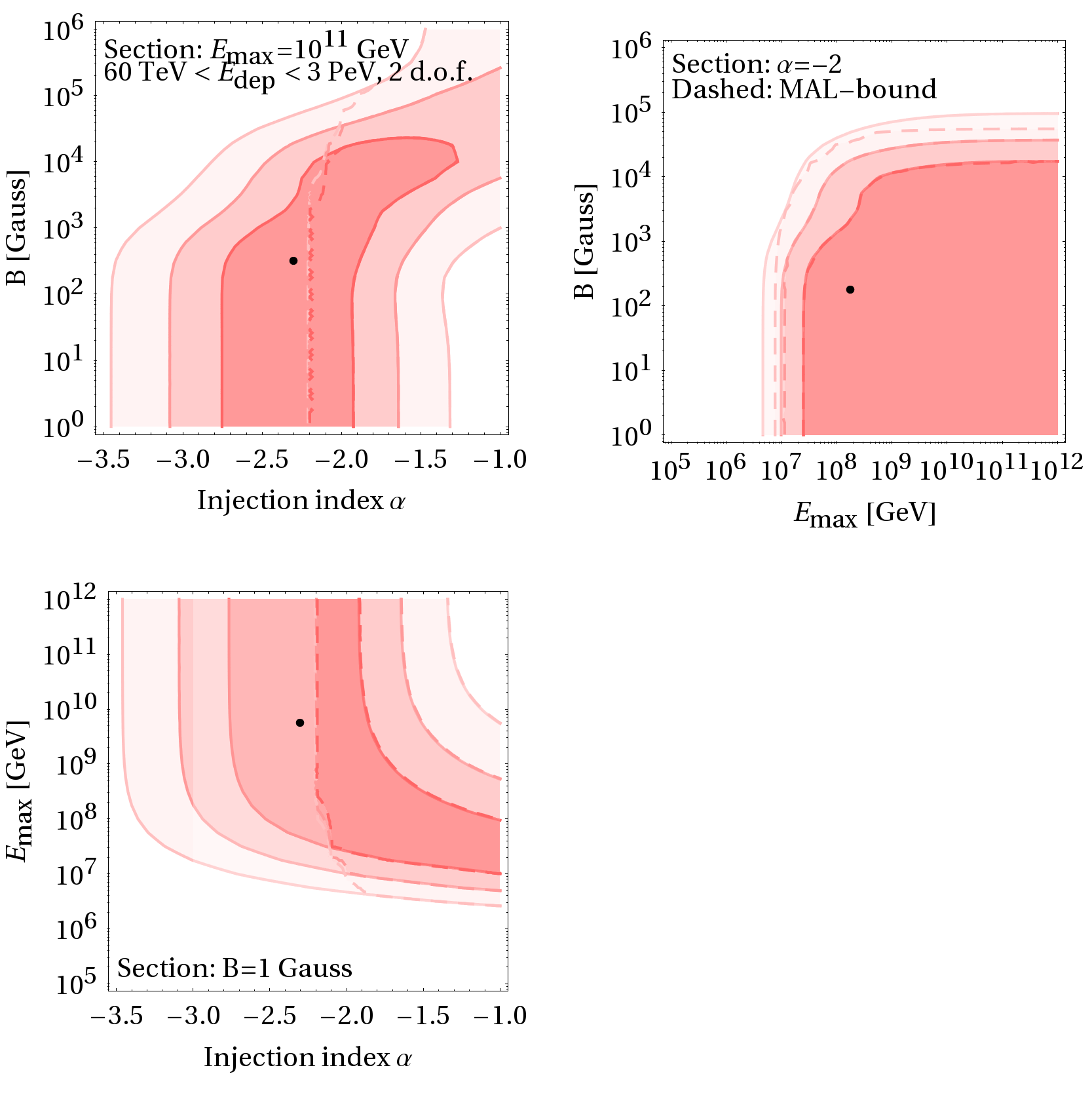}
\end{center}
\caption{\label{fig:model}  Allowed fit regions to neutrino data  at $1\sigma$, $2 \sigma$, and $3 \sigma$ (2 d.o.f.) in a three parameter ($\alpha$, $B$, $E_{\mathrm{max}}$) model for protons only, where one of the parameters is fixed in each panel (sections shown).  Here only events used by the  IceCube collaboration fit~\cite{Aartsen:2014gkd} in terms of deposited energy   $60 \, \mathrm{TeV} \le E_{\mathrm{dep}} \le 3 \, \mathrm{PeV}$ have been used. Dashed contours refer to including a generic bound by Murase, Ahlers, Lacki (MAL)~\cite{Murase:2013rfa} on the $\gamma$-ray emission from $\pi^0$s co-produced with charged pions if the spectrum  extends down to $100 \, \mathrm{GeV}$, where it has to obey the Fermi isotropic background bound~\cite{Abdo:2010nz}. The best-fit (for the solid contours) is marked by the dot.}
\end{figure*}

Here  we first reproduce the IceCube spectral fit in \Ref~\cite{Aartsen:2014gkd}, which is based on a sub-sample of the 37 events with  $60 \, \mathrm{TeV} \le E_{\mathrm{dep}} \le 3 \, \mathrm{PeV}$. All following figures will be based on the full data sample.
In \figu{model}, the allowed fit region (filled contours) are shown for in a three parameter ($\alpha$, $B$, $E_{\mathrm{max}}$) model for protons only, where one of the parameters is fixed in each panel.  Most noteworthy, for $B \le 10^2 \, \mathrm{G}$, we obtain a spectral index $\alpha = -2.3 \pm 0.3$ in consistency with the collaboration results. 
Thus, although our procedure qualitatively deviates from internal analyses of the IceCube collaboration in a few ways (\eg, mapping from deposited to reconstructed energy/energy reconstruction, details on background model, systematical errors), we can roughly reproduce their results in order to test more complicated models. For instance, one can read off from this figure that $E_{\mathrm{max}}$ and $B$ can produce a cutoff as alternative to softer spectra. For instance, strong enough $B$, a spectral index $\alpha=-2$ is allowed. Note that in the following, we will use the full set of events to make the full use of statistics.

An interesting observation was made by Murase, Ahlers, and Lacki (MAL)~\cite{Murase:2013rfa}: the production of gamma-rays from $\pi^0$ decays in the sources, which are co-produced with the charged pions, may violate the Fermi isotropic background measurements~\cite{Abdo:2010nz}. The highest energy data points are at about 100~GeV, which is significantly below the measured neutrino energies and requires some extrapolation of the spectrum. In addition, gamma-rays may come from higher energies by the initiated electromagnetic cascade. While the details are somewhat model-dependent, we compute the gamma-ray flux injected at the sources based on \Ref~\cite{Kelner:2006tc}. We add a penalty $\chi^2$, imposing that $\left. E^2 \phi_\gamma \right|_{100 \, \mathrm{GeV}} = 8^{+2}_{-\infty} \, 10^{-8} \, \mathrm{GeV cm^{-2} s^{-1} sr^{-1}}$, \ie, an upper bound. The effect of this penalty can be seen as dashed curves in \figu{model}: it leads to a lower cutoff  $\alpha \gtrsim -2.2$ ($1 \sigma$). This result is consistent with \Ref~\cite{Murase:2013rfa}, where $\alpha \gtrsim -2.18$ was found.
 Note that for the chosen star formation evolution of the sources, the main contribution will come from $z \sim 1$, where the optical depth at $100 \, \mathrm{GeV}$ is still small enough such that most gamma-rays can reach us without being attenuated in photon background fields during their propagation. However, the gamma-ray constraint can be at least partially avoided if a spectral break or lower cutoff in the  energy spectrum of the non-thermal nucleons is introduced. We therefore discuss it separately in  this study.

\begin{figure*}[t!]
\begin{center}
\includegraphics[width=0.8\textwidth]{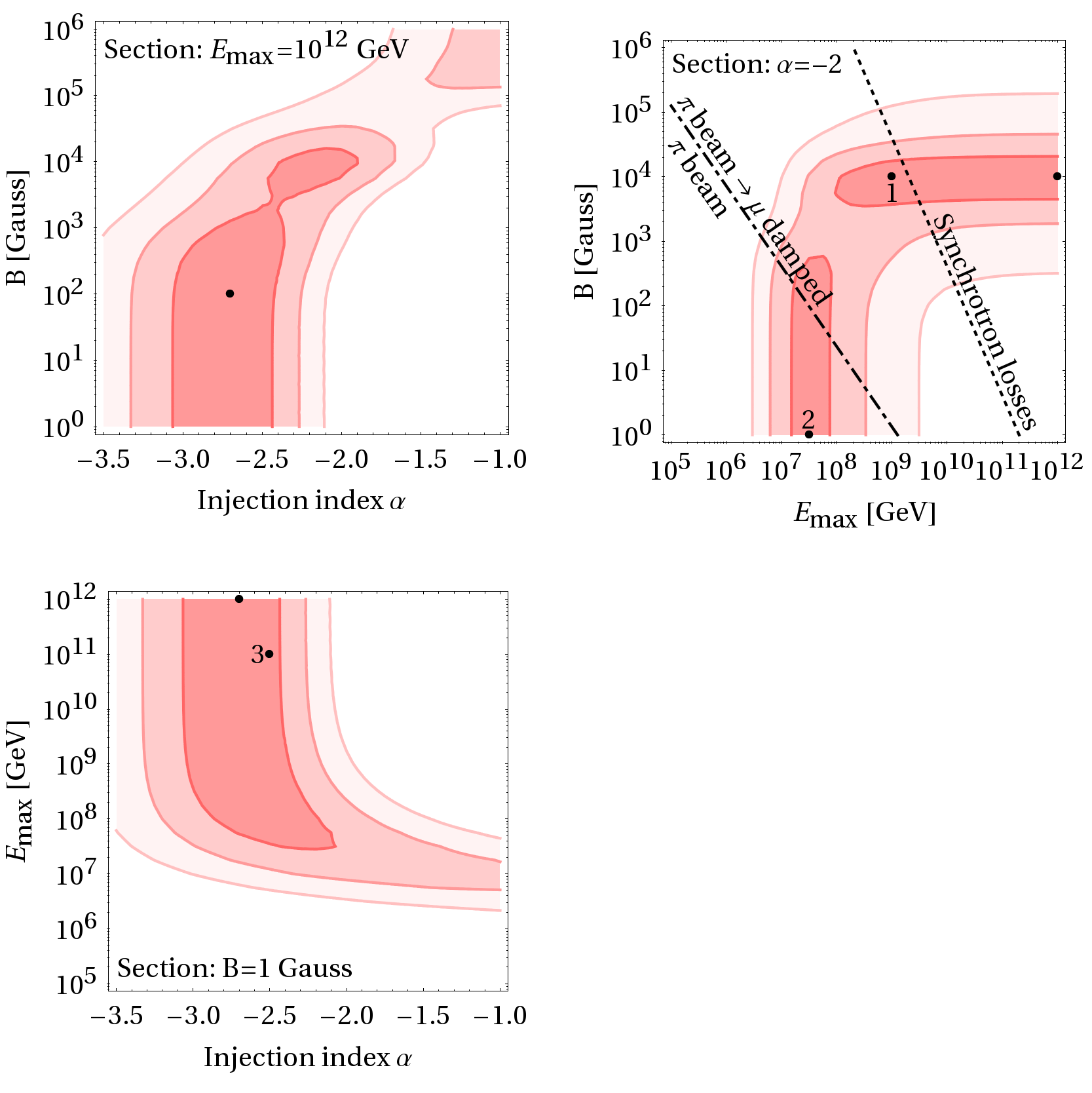}
\end{center}
\caption{\label{fig:allfit} Allowed fit regions to neutrino data  at $1\sigma$, $2 \sigma$, and $3 \sigma$ (2 d.o.f.) in a three parameter ($\alpha$, $B$, $E_{\mathrm{max}}$) model for protons only, including the full data set. The filled contours represent sections, \ie, the third (not shown) parameter in each panel is fixed to the depicted value. The overall minimum is shown as dot, and three additional test points are marked as well (see \figu{spec}). The lines in the upper right panel are discussed in the main text. }
\end{figure*}

Let us now take into account all 36~events with energy information.
The fit result is shown in \figu{allfit} for a proton composition, where one of the parameters is fixed in each panel (parameter space ``section''). The minimal $\chi^2$ is about $9$ in the left-panels, and the $\chi^2/\mathrm{d.o.f.}$ is about two. This relatively large value comes from the above mentioned tension between muon track prediction and observation, and can in principle be avoided using a different background model. However, given the small number of bins (eight), it should not be over-emphasized. Comparing to \figu{model}, we note that the spectral index $\alpha$ shifts to softer values ($\alpha=-2.7 \pm 0.2$, $1\sigma$ for 1 d.o.f.), see upper left panel for small $B$. Apart from a better matching  at low energies, the information beyond $3 \, \mathrm{PeV}$  leads to stronger constraints because no neutrinos have been seen there. However, while there is a tendency towards softer spectra $\alpha \ll -2.3$ (for the best-fit) in all tested cases, the exact best-fit value of  $\alpha$ depends on details of how the (steep) atmospheric backgrounds are implemented.
We also have a clear limit in the $E_{\mathrm{max}}$-$B$ plane (upper right panel), where the lower right corner is excluded because it would produce too many high-E events. In that panel, two distinctive regions appear at the $1 \sigma$ confidence level: one can either produce the cutoff with $10^7 \, \mathrm{GeV} \lesssim E_{\mathrm{max}} \lesssim 10^8 \, \mathrm{GeV}$, or with $B \sim 10^4 \, \mathrm{G}$. It is noteworthy that these regions can be potentially discriminated by the flavor composition of the neutrinos: Roughly on the r.h.s. of the dashed-dotted line, the neutrino production will be dominated by pion decays at the highest energies, whereas the muons lose energy faster than they decay (``muon damped source''). As a consequence, only muon neutrinos and antineutrinos will be produced at the source, which leads to a deviation from the canonical $(\nu_e:\nu_\mu:\nu_\tau) \sim (1:1:1)$ flavor composition at the detector including flavor mixing~\cite{Learned:1994wg}; see \Ref~\cite{Winter:2012xq} for a review. 

A potential theoretical constraint comes from proton synchrotron losses: not necessarily all of the regions shown in \figu{allfit} can be reached, as protons may lose energy in magnetic fields faster than they can be accelerated. This can be quantified using an acceleration rate for shock acceleration~\cite{Hillas:1985is}
\begin{equation}
 t^{-1}_{\mathrm{acc}}=\eta \frac{c^2 Z e B}{E} \, , \label{equ:acc}
\end{equation}
which corresponds to a constant fractional energy gain per cycle $\eta$. For efficient acceleration, one typically assumes $\eta \simeq 1$. Synchrotron losses are governed by
\begin{equation}
 t^{-1}_{\mathrm{synchr}} = \frac{Z^4 e^4 B^2 E}{9 \pi \varepsilon_0 m^4 c^5} \, , \label{equ:synchr}
\end{equation}
which means that they take over at high enough energies. Equating \equ{acc} with \equ{synchr}, one obtains
\begin{equation}
 E_{\mathrm{max}} \propto \frac{m^2}{Z^{3/2}} \sqrt{\frac{\eta}{B}}  \propto \sqrt{A} \, \sqrt{\frac{\eta}{B}} \, , \label{equ:emaxsynchr}
\end{equation}
where in the latter step  $m \propto A \propto Z$ was assumed, which is a good approximation for elements heavier than hydrogen. For hydrogen and $\eta=1$, the region where synchrotron losses dominate is on the r.h.s. of the dashed line in \figu{allfit}, upper right panel. That is, for shock acceleration and and moderately efficient acceleration, that region cannot be reached. We will consider the impact of this theoretical constraint on the fit below. Note that we do not assume relativistic boosting here. For example for gamma-ray bursts, $E_{\mathrm{max}} \simeq 10^9 \, \mathrm{GeV}$ can be reached for $100 \, \mathrm{kG}$ in the shock rest frame, which translates into $E_{\mathrm{max}} \sim 10^{11} \, \mathrm{GeV}$ in the observer's frame. Note that an additional constraint comes from the size of the acceleration region (Hillas criterium), which is however difficult to interpret for highly relativistic sources due to relativistic length contraction and energy boosting. We do not explicitly discuss this constraint here, as it involves another parameter (size of the acceleration region) which can be interesting for the interpretation of the signal in terms of specific source classes, but only adds limited new information to our generic fit.

\begin{figure}[t!]
\begin{center}
\includegraphics[width=0.95\columnwidth]{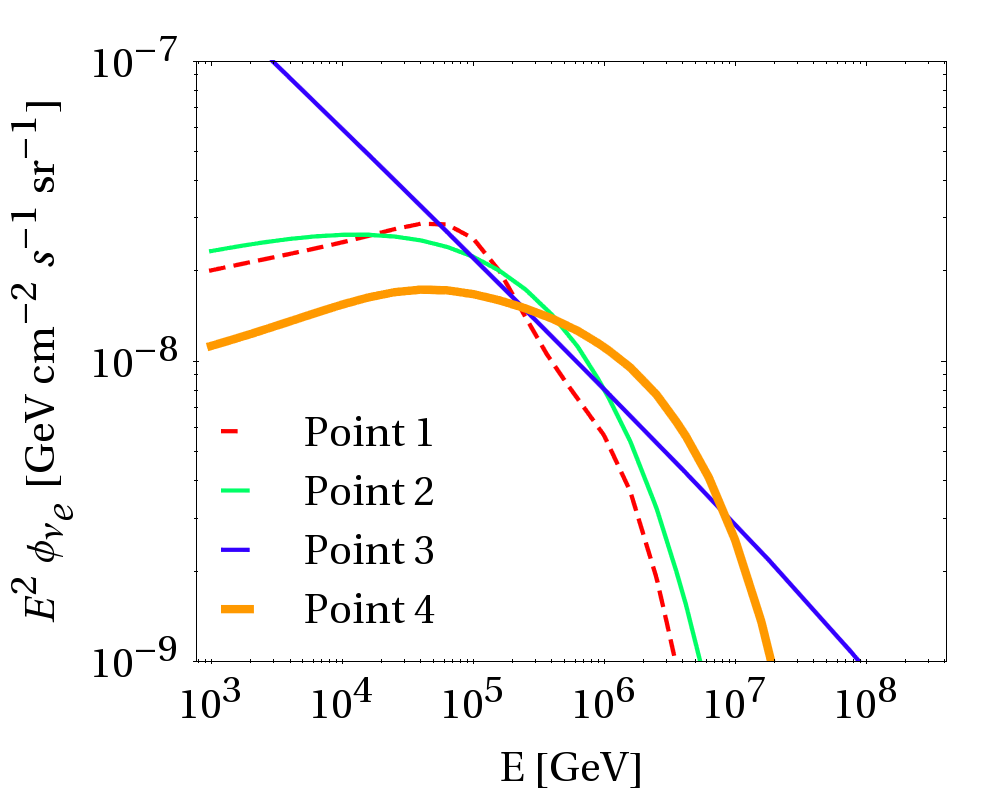} \\
\hspace*{0.2cm}\includegraphics[width=0.95\columnwidth]{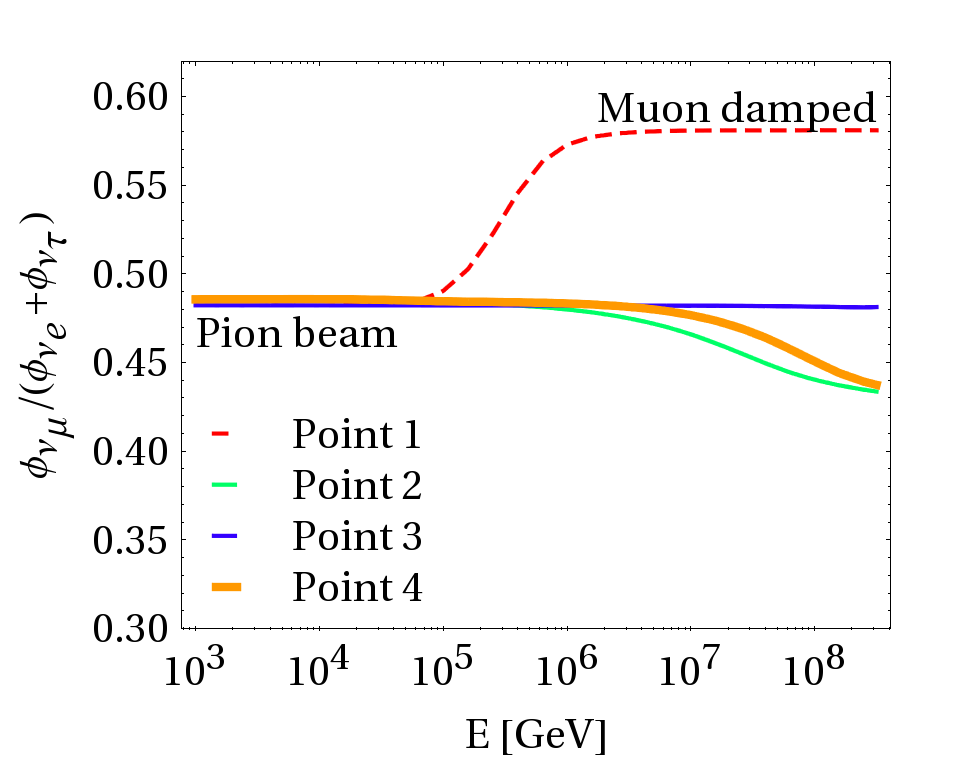}
\end{center}
\caption{\label{fig:spec} Spectra (upper panel) and flavor composition at detector (lower panel) corresponding to the points~1 to~3 marked in \figu{allfit}, where the normalization represents the best-fit to IceCube data. Point~1 refers to $\alpha=-2$, $E_{\mathrm{max}}=10^9 \, \mathrm{GeV}$, and $B=10^4 \, \mathrm{G}$, point~2 to $\alpha=-2$, $E_{\mathrm{max}}=10^{7.5} \, \mathrm{GeV}$, and $B \lesssim 1 \, \mathrm{G}$, and point~3  to $\alpha=-2.5$, $E_{\mathrm{max}}=10^{11} \, \mathrm{GeV}$, and $B \lesssim 1 \, \mathrm{G}$. Point~4 refers to the heavier composition model $\alpha=-2$, $E_{\mathrm{max}}=10^{10.1} \, \mathrm{GeV}$, $B \lesssim 1 \, \mathrm{G}$, and $\beta=0.4$, see \figu{comp}.
}
\end{figure}

So what kind of options to we have to describe the data? In order to illustrate that, three test points in the $1\sigma$ region are marked in \figu{allfit}. We show the obtained best-fit spectra for these test points and electron neutrinos in \figu{spec}, upper panel. Furthermore, we show the  flavor ratio of muon to electron and tau neutrinos at the detector, which corresponds to the ratio between induced muon tracks and cascades (without efficiencies), in the lower panel. We can identify three options:
\begin{description}
 \item[Point 1.] High $E_{\mathrm{max}}$ are allowed together with strong magnetic fields. The magnetic fields lead to a cutoff and some characteristic wiggles in the spectrum, which come together with a change of the flavor composition at PeV energies from pion beam to muon damped source, see lower panel of \figu{spec}.
\item[Point 2.] For small $B$, the cutoff can be achieved by an appropriate maximal proton energy, as discussed above. 
\item[Point 3.] Alternatively, a soft enough spectrum can describe data for small $B$ and large $E_{\mathrm{max}}$; see also \Ref~\cite{Anchordoqui:2013qsi}.
\end{description}
We will discuss Point~4 in the next section. Note that in all cases, the normalization (which is a result of the fit) is about $1.5 \, 10^{-8} \, \mathrm{GeV \, cm^{-2} \, s^{-1} \, sr^{-1}}$ at $30 \, \mathrm{TeV}$.

It is, of course, interesting to discuss what this information tells us about the sources. Point~1 corresponds to sources with strong magnetic fields, such as low luminosity gamma-ray bursts~\cite{Murase:2008mr}, ``chocked'' gamma-ray bursts~\cite{Razzaque:2004yv,Ando:2005xi,Razzaque:2005bh}, or (extra-galactic) micoquasars or pulsars. Point~2 may correspond to starburst galaxies~\cite{Loeb:2006tw}, galaxy clusters/groups~\cite{Murase:2013rfa} or radio galaxies~\cite{Tjus:2014dna}. And Point~3 may come from any extra-galactic population, where the main challenge is to accommodate $\alpha \ll -2.2$ with the theory of Fermi acceleration. On possibility is the effect of turbulence on Fermi shock acceleration which may cause such effects~\cite{Lemoine:2006gg}, another one is that the overall spectral index comes from convoluting a harder spectrum with an appropriate luminosity distribution function~\cite{Kachelriess:2005xh}. 

Maybe even more interesting is the conceptual question if these neutrinos can come from the sources of the UHECRs. Considering Point~1, which is taking into account the synchrotron loss constraint in \figu{allfit}, the maximal energy can only be high enough to reach the UHECR range $E > 10^{10} \, \mathrm{GeV}$ if large Lorentz boosts are involved. Point~2, on the other hand, cannot be accommodated with the UHECR paradigm, because $E_{\mathrm{max}}$ is too low. For Point~3, the discussion is much more complicated: While it can be in principle accommodated with the UHECR paradigm, the soft spectrum tends to lead to neutrino overproduction at PeV energies if one normalizes the UHECR range to normalization. This is discussed for gamma-ray bursts in \Refs~\cite{Ahlers:2011jj,Baerwald:2014zga}, and, in a more generic context, in \Ref~\cite{Katz:2013ooa}. A model-independent ``proof'' seems, however, more difficult.

\begin{figure*}[t!]
\begin{center}
\includegraphics[width=0.8\textwidth]{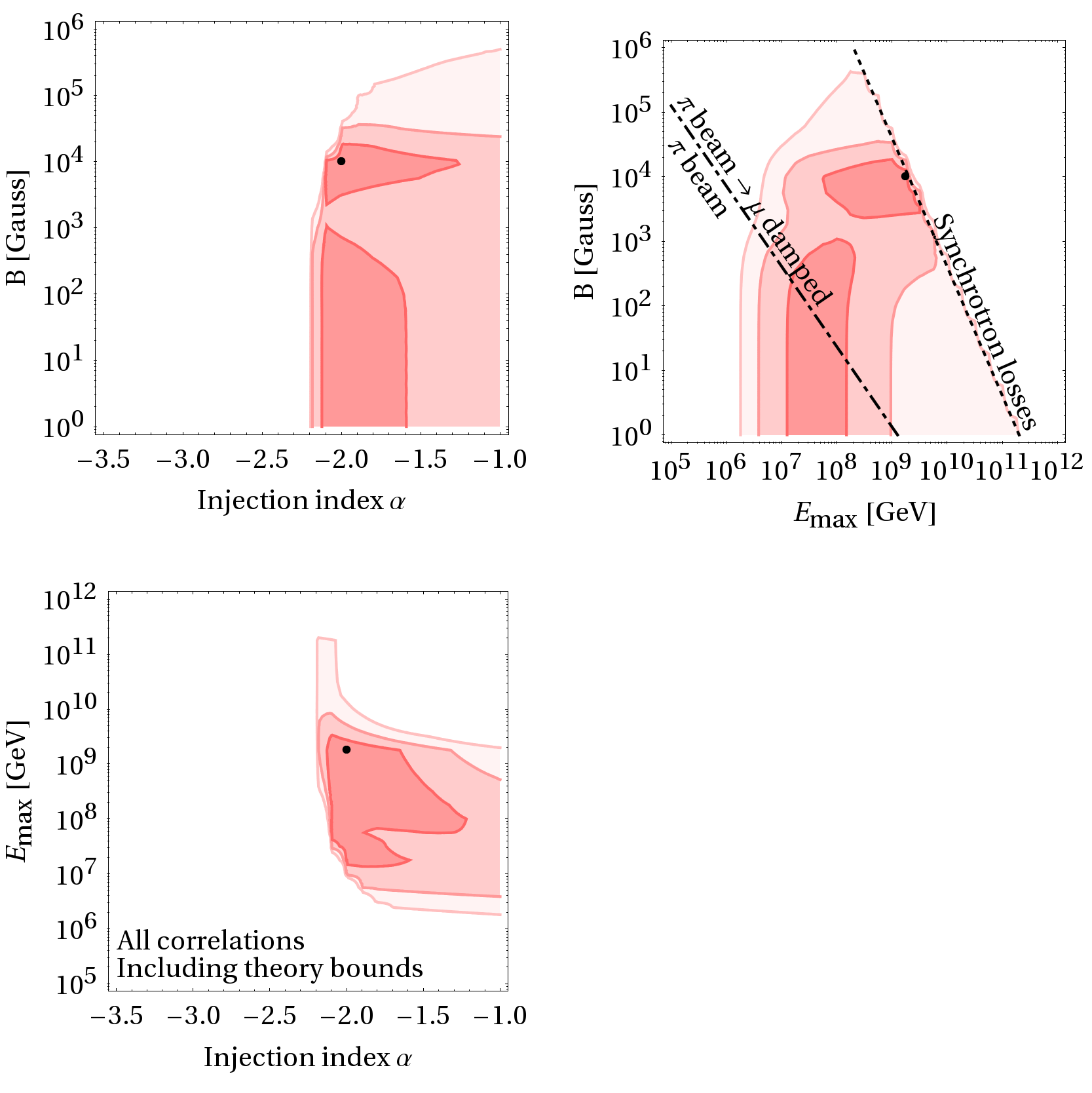}
\end{center}
\caption{\label{fig:allfit_th} Allowed fit regions to neutrino data  at $1\sigma$, $2 \sigma$, and $3 \sigma$ (2 d.o.f.) in a three parameter ($\alpha$, $B$, $E_{\mathrm{max}}$) model for protons only. Here the full parameter degeneracy is taken into account, \ie, the $\chi^2$ is minimized over the third parameter in each panel. Here theoretical exclusion limits are included as $\chi^2$ penalties: the Murase-Ahlers-Lacki bound on $\gamma$-ray observations, and the region unreachable because of synchrotron losses dominating the maximal proton energy as a cutoff.}
\end{figure*}

In \figu{allfit_th}, we take into account multi-parameter correlations. That is, we show the projections including the minimization of the parameter not shown in each panel. This increases the fit region regions dramatically. On the other hand, we include the Murase-Ahlers-Lacki bound, which leads to $\alpha \gtrsim -2.2$, and the synchrotron loss constraint, explicitly shown in the upper right panel. These theoretical constraints reduce the size of the fit regions. The main result, which can be read off from the left panels, is that spectral indices compatible with Fermi acceleration are preferred in combination with a cutoff of the maximal proton energy, whereas a wide range of magnetic fields are possible. The best-fit at $\alpha \simeq 2$ and $E_{\mathrm{max}} \simeq 2 \, 10^9 \, \mathrm{GeV}$ points towards conventional scenarios of Fermi shock acceleration together with a maximal proton energy cutoff compatible with the ankle of the cosmic ray spectrum observed in our Galaxy. It is therefore plausible that the neutrinos are produced under similar conditions, such as in starburst galaxies.

\section{Nuclei-matter interactions, and the UHECR paradigm}

Adding the composition to the parameters of the model increases the complexity to a level such that no meaningful information can be obtained from current data due to limited statistics. We therefore focus on the key issue, the proton composition has not been convincingly successful to describe: can the potential cutoff at PeV energies be reconciled with the UHECR paradigm, taking into account that the composition could be as heavy as iron at the highest energies~\cite{Abraham:2010yv}? 

Let us assume that the  magnetic fields are small enough such that magnetic field effects on the secondaries are negligible. That is a necessary condition such that the maximal energies are not suppressed by synchrotron losses in our standard scenario (apply, for instance, \equ{emaxsynchr} to dashed curve in \figu{allfit} for $E_{\mathrm{max}}=10^{11} \, \mathrm{GeV}$). Let us furthermore assume that $\alpha \simeq -2$ in consistency with the argument in \Ref~\cite{Katz:2013ooa}, and that $A(E_{\mathrm{max}})=56$ (iron). We then assume that the maximal energy is element-dependent, 
which leads to an energy-dependent composition  parameterized as
\begin{equation}
A(E) = \mathrm{max} \left( 1, \, 56 \times \left(\frac{E}{E_{\mathrm{max}}} \right)^\beta \right) \, ,
\label{equ:a}
\end{equation}
\ie, $A(E) \ge 1$ and $A(E_{\mathrm{max}})=56$. The coefficient $\beta$ describes how $A$ scales with energy, and we treat it as a continuous parameter.
It is, however, useful to consider a few examples:  
\begin{description}
\item[Rigidity scaling.] This is the most often used approach, also known as necessary condition formulated by Hillas~\cite{Hillas:1985is}: the Larmor radius has to be smaller than the acceleration region. This can be also described using \equ{acc} by $t^{-1}_{\mathrm{acc}}=t^{-1}_{\mathrm{lim}} \simeq c/R$, where $R$ is the size of the region and $t^{-1}_{\mathrm{lim}}$ the limiting timescale determined by the size of the region (typically the dynamical timescale, escape timescale, or adiabatic cooling timescale). As a consequence, $E/Z$ (the rigidity) is constant for constant $B$ and $R$, which means that higher energies can be reach for higher charges. Since $Z \sim A/2$, one has $\beta \simeq 1$.
\item[Synchrotron-loss dominated $\boldsymbol{E_{\mathrm{max}}}$.] From \equ{emaxsynchr} describing a shock acceleration scenario, we can immediately read off that $A \propto E^2$, \ie, $\beta=2$, if the maximal energy is limited by synchrotron losses.
\end{description}
It is generically difficult to obtain coefficients $\beta < 1$ unless the timescale constraining the maximal energy slightly drops with energy and does not scale with rigidity. This may be achieved in scenarios where \eg\ photo-disintegration dominates the highest energies~\cite{Murase:2008mr}.  However, note that the composition has been observed to be light at $10^9 \, \mathrm{GeV}$~\cite{Abbasi:2009nf,Abraham:2010yv}, which gives a constraint $E_{\mathrm{max}}(\beta)$ which can be easily derived from \equ{a} using $A(10^9 \, \mathrm{GeV})=1$. 

\begin{figure}[tp]
\begin{center}
\includegraphics[width=0.8\columnwidth]{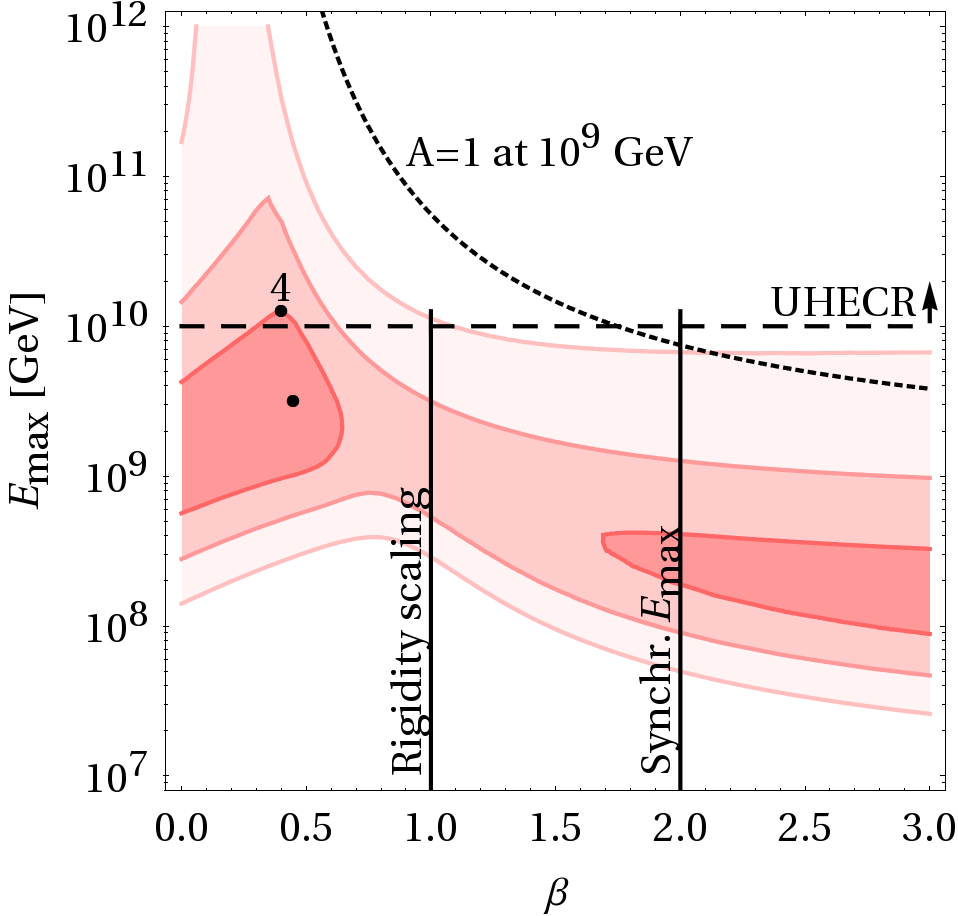}
\end{center}
\caption{\label{fig:comp}  Allowed fit region to neutrino data at $1\sigma$, $2 \sigma$, and $3 \sigma$ (2 d.o.f., filled contours) as a function of $\beta$ and $E_{\mathrm{max}}$ for $\alpha=-2$, and $B$ small enough such that magnetic field effects on the secondaries can be neglected. Here the composition is chosen to be iron at the highest energy $E_{\mathrm{max}}$, and the composition is assumed to be energy dependent with $A(E) = \mathrm{max}(1, \, 56 \times (E/E_\mathrm{max})^\beta)$. The vertical lines correspond to different acceleration scenarios, as discussed in the main text.  If the composition is to be dominated by protons at $10^9 \, \mathrm{GeV}$, the dotted curve has to be matched.
}
\end{figure}

The current best-fit region in terms of $\beta$ and $E_{\mathrm{max}}$ is shown in \figu{comp} for fixed  $\alpha=-2$. It is clear from the figure that $E_{\mathrm{max}} \gg 10^{10} \, \mathrm{GeV}$, required to describe UHECR observations, implies that $\beta<1$. Extremely high energies are allowed for $0.05 \lesssim \beta \lesssim 0.35$, which requires unconventional assumptions for the acceleration-radiation scenarios -- as discussed above. This scenario neither requires strong enough magnetic fields, nor a spectral index softer than $\alpha=-2$; the cutoff is instead produced by a change of the composition.
Note that  neutrino data can in that case be used to infer on the acceleration of the heavier elements itself, or to model the injected UHECR composition from the sources in propagation codes. 
The spectrum corresponding to test point~4 is also shown in \figu{spec} (upper panel). It peaks at somewhat higher energies than the other spectra. It is probably noteworthy that, because of $\alpha=-2$, there are no issues with the MAL-bound in this case, as we have explicitly tested.  However, this scenario is in slight tension with the observed light composition at $10^9 \, \mathrm{GeV}$, which is for protons (extreme case) shown as dotted curve in \figu{comp}. If slightly heavier compositions are admitted at $10^9 \, \mathrm{GeV}$, such as helium, the dotted curve moves closer to the fit contours.

\section{Future expectations for IceCube performance}

\begin{figure}[t]
\begin{center}
\includegraphics[width=0.8\columnwidth]{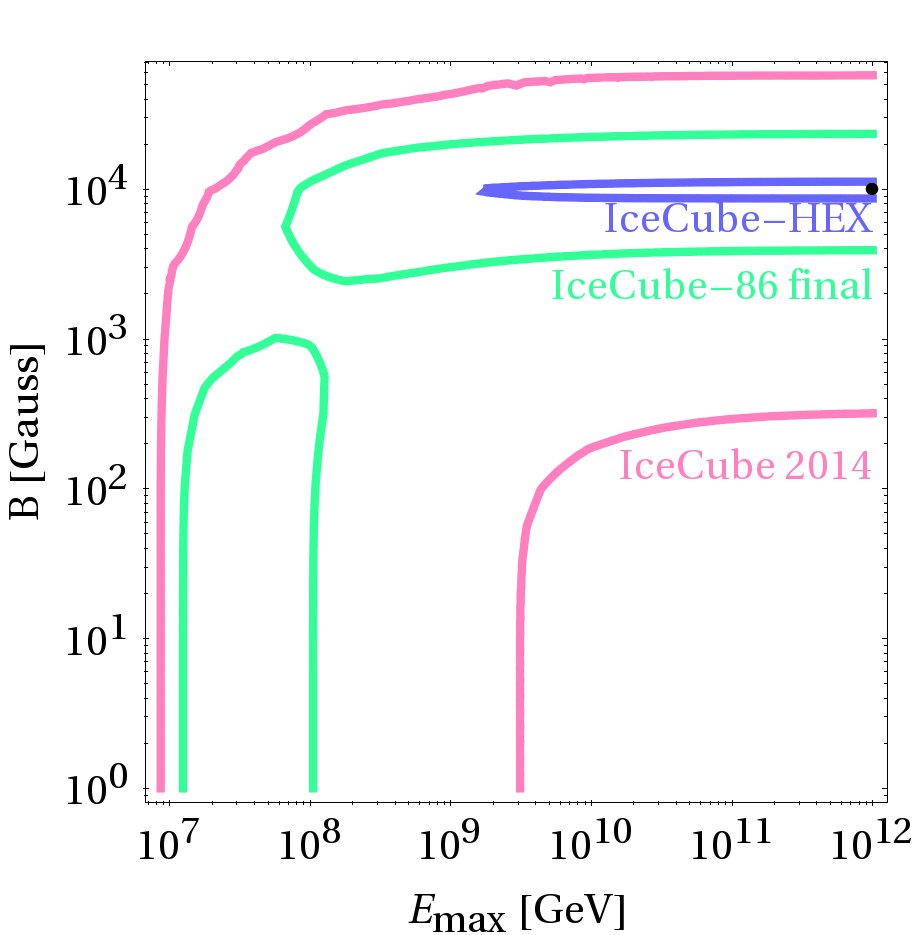}
\end{center}
\caption{\label{fig:extra} Extrapolation of $3 \sigma$-allowed region as a function of $E_{\mathrm{max}}$ and $B$ (section for protons and $\alpha=-2$, corresponding to \figu{allfit}, upper right panel, but including the MAL bound). ``IceCube 2014'' corresponds to current data (37 events), ``IceCube-86 final'' to four times the current exposure, and ``IceCube-HEX'' to a possible high-energy extension with fourty times the current exposure (factor ten larger mass operated over about a decade).  The best-fit is marked by a dot.
}
\end{figure}

We finally discuss what can be learned from future upgrades of IceCube. As an example, let us choose the the $3 \sigma$-allowed region as a function of $E_{\mathrm{max}}$ and $B$ corresponding to \figu{allfit}, upper right panel (but including the MAL bound). The outer (red) curve shows the constraint from current data, the middle green region the expected result from IceCube-86 over about a decade (four times current exposure), the blue region the expected result from a future high-energy extension (HEX) with about ten times the size of IceCube, operated over about a decade. This figure illustrates that while better information will be available in a few years from now with upcoming data, high precision will require an upgrade of IceCube. In this particular case IceCube-86 cannot discriminate between a cutoff from magnetic field effects or a cutoff in the proton spectrum, corresponding to test points~1 and~2. However, the precision in IceCube-HEX will even allow to exploit the transition in the flavor composition expected by magnetic field effects, and discriminate between these regions. This is evident from the event rates. Consider \eg\ Point~1 and the muon track and cascade bins between 1 and 2~PeV. For current statistics, the expectations are 0.6 tracks and 0.9 cascades for these bins, whereas for IceCube-HEX, the expectations are 24 muon tracks and 37 cascades.
This leads to a statistical relative error of about $1/\sqrt{24} \simeq 20\%$
 The muon track to shower ratio is about 20\% increased in the muon damped case (coinciding with this energy range for the chosen test point) compared to the pion beam case after flavor mixing, see \figu{spec} (lower panel, between 1~and 2~PeV), which means that the statistics between muon tracks and showers becomes meaningful. This is a significant qualitative advance compared to the full statistics IceCube-86 analysis.

\section{Summary and discussion}

We have studied the interpretation of IceCube data in the production scenario of nuclei-matter interactions, for which the neutrino spectrum follows the non-thermal spectrum of the nuclei. Compared to earlier studies, we have taken into account possible magnetic field effects on the secondary muons and pions and the composition of the accelerated nuclei. We have especially focused on the reproduction of the spectrum, where the flavor composition (cascades versus muon tracks) has been implied as well. We have essentially identified four different options for the initial spectrum of the protons/nuclei to reproduce current data, which can all avoid the overproduction of events beyond a few PeV:
\begin{enumerate}
 \item
  An unbroken  power law with $\alpha \sim -2$ and magnetic fields $B \sim 10^4 \, \mathrm{G}$ in the source, leading to magnetic field effects on the secondary muons and pions. This may be realized in certain populations of gamma-ray bursts or (extra-galactic) micoquasars or pulsars. 
 \item
  A power law with $\alpha \sim -2$ with a break or cutoff about $10^7$ to $10^8 \, \mathrm{GeV}$ (for protons). This option may be favorable for models of starburst galaxies, galaxy clusters/groups,  or radio galaxies.
 \item 
  An unbroken power law significantly softer than $E^{-2}$, where we find best-fit values for $\alpha$ between $-2.7$ and $-2.3$ depending on analysis range and atmospheric background model. This option is the simplest possible one, also mentioned in the recent IceCube three-year analysis~\cite{Aartsen:2014gkd}, where we typiclly find softer indices because of the larger analysis energy range. A major drawback has been pointed out by Murase, Ahlers, and Lacki (MAL)~\cite{Murase:2013rfa}: the spectrum consequently exceeds the isotropic gamma-ray background at lower energies, unless a low-energy break is introduced.
 \item
  An unbroken power law  with $\alpha \sim -2$ and a flat enough change of the composition of the non-thermal spectra in the source from lighter to heavier elements at the highest energies, which can be potentially as high as $10^{12} \, \mathrm{GeV}$.
\end{enumerate}
While options~1 to~3 have been identified in similar forms in the literature, option~4 is entirely new. Note that regarding options~1 and~2, similar possibilities are obtained for photohadronic interactions producing neutrinos, for which the spectral shape is determined by both the nuclei and target photon spectra; see \Ref~\cite{Winter:2013cla}. 

We have been especially interested in which of the above options can be reconciled with the question that the observed neutrinos stem from the sources of the UHECRs, see also discussion in \Refs~\cite{Liu:2013wia,Kistler:2013my}.  Option 2) is obviously incompatible with this assumption if the break or cutoff can be also found in the escaping proton spectrum injected into the intergalatic medium. Exceptions could be scenarios with energy-dependent diffusive escape, as in that case the steady proton spectrum (responsible for neutrino production) can have a break, whereas the escaping proton spectrum can be different; see \eg\ \Ref~\cite{Liu:2013wia} for hypernova remnants. Option 1) faces the problem that synchrotron losses typically limit the maximal proton energy, and do not allow for high enough maximal energies (unless strong Lorentz boosts are involved, such as in gamma-ray bursts). Option 3), on the other hand, tends to lead to an overproduction of neutrinos at PeV energies exceeding the current IceCube observations when normalized to the UHECR observations, see \Ref~\cite{Katz:2013ooa} for a generic discussion -- while it also violates the MAL bound. We have therefore identified option 4) as the most promising one, especially in the light of recent Auger observation pointing towards a heavier composition at the highest energies. It is generically compatible with the UHECR paradigm, as the acceleration energies can be high enough to describe observations, and it is also compatible with the MAL bound. In fact, we have demonstrated that neutrino data can be used to test the acceleration mechanism in that scenario. The result from neutrino observations may, on the other hand, serve as an input for the UHECR injection at the sources in UHECR propagation models. However, the scenario is in slight tension with a very light composition at $10^9 \, \mathrm{GeV}$.

Finally, we have pointed out that future precision measurements will require significant volume upgrades, such as an high-energy extension of IceCube or KM3NeT in the Mediterranean. While this statement is most certainly generically true, we have shown that the current IceCube experiment will not be sufficient to exploit the flavor information from muon tracks versus cascades in a statistically meaningful manner to discriminate between pion beam and muon damped sources. The reason is that the flavor transition is expected at high enough energies to describe the cutoff at PeV energies, where the event rates are very low. A significantly larger (about a factor of ten) volume upgrade would, however, allow for a discrimination between options 1) and 2), where the flavor information will be {\bf the} qualitatively new ingredient. It is therefore important to optimize an upgrade in a way to preserve the flavor or topology identification capability.

\subsection*{Acknowledgments}

 I would like to thank Markus Ackermann, Sergio Palomares-Ruiz, Karl Mannheim, Lars Mohrmann, Julia Tjus, Xiang-Yu Wang, and Nathan Whitehorn for useful discussions and comments on aspects of this work, and Lars Mohrmann for reading the manuscript.

I would like to acknowledge support from DFG grants WI 2639/3-1 and WI 2639/4-1,  the ``Helmholtz Alliance for Astroparticle Physics HAP'',  funded by the Initiative and Networking fund of the Helmholtz
association. I would also like to acknowledge support from the  Nordita program ``News in neutrino physics'' from April 21-May 2, 2014, where parts of this work were carried out. 

\vspace*{0.5cm}


\begin{thebibliography}{54}
\expandafter\ifx\csname natexlab\endcsname\relax\def\natexlab#1{#1}\fi
\expandafter\ifx\csname bibnamefont\endcsname\relax
  \def\bibnamefont#1{#1}\fi
\expandafter\ifx\csname bibfnamefont\endcsname\relax
  \def\bibfnamefont#1{#1}\fi
\expandafter\ifx\csname citenamefont\endcsname\relax
  \def\citenamefont#1{#1}\fi
\expandafter\ifx\csname url\endcsname\relax
  \def\url#1{\texttt{#1}}\fi
\expandafter\ifx\csname urlprefix\endcsname\relax\def\urlprefix{URL }\fi
\providecommand{\bibinfo}[2]{#2}
\providecommand{\eprint}[2][]{\url{#2}}

\bibitem[{\citenamefont{Aartsen et~al.}(2013{\natexlab{a}})}]{Aartsen:2013bka}
\bibinfo{author}{\bibfnamefont{M.}~\bibnamefont{Aartsen}} \bibnamefont{et~al.}
  (\bibinfo{collaboration}{IceCube Collaboration}),
  \bibinfo{journal}{Phys. Rev. Lett.} \textbf{\bibinfo{volume}{111}},
  \bibinfo{pages}{021103} (\bibinfo{year}{2013}{\natexlab{a}}),
  \eprint{1304.5356}.

\bibitem[{\citenamefont{Aartsen et~al.}(2013{\natexlab{b}})}]{Aartsen:2013jdh}
\bibinfo{author}{\bibfnamefont{M.}~\bibnamefont{Aartsen}} \bibnamefont{et~al.}
  (\bibinfo{collaboration}{IceCube}), \bibinfo{journal}{Science}
  \textbf{\bibinfo{volume}{342}}, \bibinfo{pages}{1242856}
  (\bibinfo{year}{2013}{\natexlab{b}}), \eprint{1311.5238}.

\bibitem[{\citenamefont{Aartsen et~al.}(2014)}]{Aartsen:2014gkd}
\bibinfo{author}{\bibfnamefont{M.}~\bibnamefont{Aartsen}} \bibnamefont{et~al.}
  (\bibinfo{collaboration}{IceCube Collaboration}),
  \bibinfo{journal}{Phys. Rev. Lett.} \textbf{\bibinfo{volume}{113}},
  \bibinfo{pages}{101101} (\bibinfo{year}{2014}), \eprint{1405.5303}.

\bibitem[{\citenamefont{Ahlers and Halzen}(2014)}]{Ahlers:2014ioa}
\bibinfo{author}{\bibfnamefont{M.}~\bibnamefont{Ahlers}} \bibnamefont{and}
  \bibinfo{author}{\bibfnamefont{F.}~\bibnamefont{Halzen}}
  (\bibinfo{year}{2014}), \eprint{1406.2160}.

\bibitem[{\citenamefont{Katz}(2006)}]{Katz:2006wv}
\bibinfo{author}{\bibfnamefont{U.~F.} \bibnamefont{Katz}},
  \bibinfo{journal}{Nucl. Instrum. Meth.} \textbf{\bibinfo{volume}{A567}},
  \bibinfo{pages}{457} (\bibinfo{year}{2006}), \eprint{astro-ph/0606068}.

\bibitem[{\citenamefont{Anchordoqui
  et~al.}(2014{\natexlab{a}})\citenamefont{Anchordoqui, Barger, Cholis,
  Goldberg, Hooper et~al.}}]{Anchordoqui:2013dnh}
\bibinfo{author}{\bibfnamefont{L.~A.} \bibnamefont{Anchordoqui}},
  \bibinfo{author}{\bibfnamefont{V.}~\bibnamefont{Barger}},
  \bibinfo{author}{\bibfnamefont{I.}~\bibnamefont{Cholis}},
  \bibinfo{author}{\bibfnamefont{H.}~\bibnamefont{Goldberg}},
  \bibinfo{author}{\bibfnamefont{D.}~\bibnamefont{Hooper}},
  \bibnamefont{et~al.}, \bibinfo{journal}{Journal of High Energy Astrophysics}
  \textbf{\bibinfo{volume}{1-2}}, \bibinfo{pages}{1}
  (\bibinfo{year}{2014}{\natexlab{a}}), \eprint{1312.6587}.

\bibitem[{\citenamefont{Winter}(2013)}]{Winter:2013cla}
\bibinfo{author}{\bibfnamefont{W.}~\bibnamefont{Winter}},
  \bibinfo{journal}{Phys. Rev.} \textbf{\bibinfo{volume}{D88}},
  \bibinfo{pages}{083007} (\bibinfo{year}{2013}), \eprint{1307.2793}.

\bibitem[{\citenamefont{Kachelriess et~al.}(2008)\citenamefont{Kachelriess,
  Ostapchenko, and Tomas}}]{Kachelriess:2007tr}
\bibinfo{author}{\bibfnamefont{M.}~\bibnamefont{Kachelriess}},
  \bibinfo{author}{\bibfnamefont{S.}~\bibnamefont{Ostapchenko}},
  \bibnamefont{and} \bibinfo{author}{\bibfnamefont{R.}~\bibnamefont{Tomas}},
  \bibinfo{journal}{Phys. Rev.} \textbf{\bibinfo{volume}{D77}},
  \bibinfo{pages}{023007} (\bibinfo{year}{2008}), \eprint{0708.3047}.

\bibitem[{\citenamefont{Lipari et~al.}(2007)\citenamefont{Lipari, Lusignoli,
  and Meloni}}]{Lipari:2007su}
\bibinfo{author}{\bibfnamefont{P.}~\bibnamefont{Lipari}},
  \bibinfo{author}{\bibfnamefont{M.}~\bibnamefont{Lusignoli}},
  \bibnamefont{and} \bibinfo{author}{\bibfnamefont{D.}~\bibnamefont{Meloni}},
  \bibinfo{journal}{Phys. Rev.} \textbf{\bibinfo{volume}{D75}},
  \bibinfo{pages}{123005} (\bibinfo{year}{2007}), \eprint{0704.0718}.

\bibitem[{\citenamefont{H{\"u}mmer et~al.}(2010)\citenamefont{H{\"u}mmer,
  Maltoni, Winter, and Yaguna}}]{Hummer:2010ai}
\bibinfo{author}{\bibfnamefont{S.}~\bibnamefont{H{\"u}mmer}},
  \bibinfo{author}{\bibfnamefont{M.}~\bibnamefont{Maltoni}},
  \bibinfo{author}{\bibfnamefont{W.}~\bibnamefont{Winter}}, \bibnamefont{and}
  \bibinfo{author}{\bibfnamefont{C.}~\bibnamefont{Yaguna}},
  \bibinfo{journal}{Astropart. Phys.} \textbf{\bibinfo{volume}{34}},
  \bibinfo{pages}{205} (\bibinfo{year}{2010}), \eprint{1007.0006}.

\bibitem[{\citenamefont{Kashti and Waxman}(2005)}]{Kashti:2005qa}
\bibinfo{author}{\bibfnamefont{T.}~\bibnamefont{Kashti}} \bibnamefont{and}
  \bibinfo{author}{\bibfnamefont{E.}~\bibnamefont{Waxman}},
  \bibinfo{journal}{Phys. Rev. Lett.} \textbf{\bibinfo{volume}{95}},
  \bibinfo{pages}{181101} (\bibinfo{year}{2005}), \eprint{astro-ph/0507599}.

\bibitem[{\citenamefont{Murase and Nagataki}(2006)}]{Murase:2005hy}
\bibinfo{author}{\bibfnamefont{K.}~\bibnamefont{Murase}} \bibnamefont{and}
  \bibinfo{author}{\bibfnamefont{S.}~\bibnamefont{Nagataki}},
  \bibinfo{journal}{Phys. Rev.} \textbf{\bibinfo{volume}{D73}},
  \bibinfo{pages}{063002} (\bibinfo{year}{2006}), \eprint{astro-ph/0512275}.

\bibitem[{\citenamefont{Baerwald et~al.}(2012)\citenamefont{Baerwald,
  H{\"u}mmer, and Winter}}]{Baerwald:2011ee}
\bibinfo{author}{\bibfnamefont{P.}~\bibnamefont{Baerwald}},
  \bibinfo{author}{\bibfnamefont{S.}~\bibnamefont{H{\"u}mmer}},
  \bibnamefont{and} \bibinfo{author}{\bibfnamefont{W.}~\bibnamefont{Winter}},
  \bibinfo{journal}{Astropart. Phys.} \textbf{\bibinfo{volume}{35}},
  \bibinfo{pages}{508} (\bibinfo{year}{2012}), \eprint{1107.5583}.

\bibitem[{\citenamefont{Reynoso and Romero}(2009)}]{Reynoso:2008gs}
\bibinfo{author}{\bibfnamefont{M.~M.} \bibnamefont{Reynoso}} \bibnamefont{and}
  \bibinfo{author}{\bibfnamefont{G.~E.} \bibnamefont{Romero}},
  \bibinfo{journal}{Astron. Astrophys.} \textbf{\bibinfo{volume}{493}},
  \bibinfo{pages}{1} (\bibinfo{year}{2009}), \eprint{0811.1383}.

\bibitem[{\citenamefont{Baerwald and Guetta}(2013)}]{Baerwald:2012yd}
\bibinfo{author}{\bibfnamefont{P.}~\bibnamefont{Baerwald}} \bibnamefont{and}
  \bibinfo{author}{\bibfnamefont{D.}~\bibnamefont{Guetta}},
  \bibinfo{journal}{Astrophys. J.} \textbf{\bibinfo{volume}{773}},
  \bibinfo{pages}{159} (\bibinfo{year}{2013}), \eprint{1212.1457}.

\bibitem[{\citenamefont{Winter}(2012)}]{Winter:2012xq}
\bibinfo{author}{\bibfnamefont{W.}~\bibnamefont{Winter}},
  \bibinfo{journal}{Adv. High Energy Phys.} \textbf{\bibinfo{volume}{2012}},
  \bibinfo{pages}{586413} (\bibinfo{year}{2012}), \eprint{1201.5462}.

\bibitem[{\citenamefont{Abraham et~al.}(2010)}]{Abraham:2010yv}
\bibinfo{author}{\bibfnamefont{J.}~\bibnamefont{Abraham}} \bibnamefont{et~al.}
  (\bibinfo{collaboration}{Pierre Auger Observatory Collaboration}),
  \bibinfo{journal}{Phys. Rev. Lett.} \textbf{\bibinfo{volume}{104}},
  \bibinfo{pages}{091101} (\bibinfo{year}{2010}), \eprint{1002.0699}.

\bibitem[{\citenamefont{Kelner et~al.}(2006)\citenamefont{Kelner, Aharonian,
  and Bugayov}}]{Kelner:2006tc}
\bibinfo{author}{\bibfnamefont{S.}~\bibnamefont{Kelner}},
  \bibinfo{author}{\bibfnamefont{F.~A.} \bibnamefont{Aharonian}},
  \bibnamefont{and} \bibinfo{author}{\bibfnamefont{V.}~\bibnamefont{Bugayov}},
  \bibinfo{journal}{Phys. Rev.} \textbf{\bibinfo{volume}{D74}},
  \bibinfo{pages}{034018} (\bibinfo{year}{2006}), \eprint{astro-ph/0606058}.

\bibitem[{\citenamefont{Joshi et~al.}(2014)\citenamefont{Joshi, Winter, and
  Gupta}}]{Joshi:2013aua}
\bibinfo{author}{\bibfnamefont{J.~C.} \bibnamefont{Joshi}},
  \bibinfo{author}{\bibfnamefont{W.}~\bibnamefont{Winter}}, \bibnamefont{and}
  \bibinfo{author}{\bibfnamefont{N.}~\bibnamefont{Gupta}},
  \bibinfo{journal}{MNRAS} \textbf{\bibinfo{volume}{439}},
  \bibinfo{pages}{3414} (\bibinfo{year}{2014}), \eprint{1310.5123}.

\bibitem[{\citenamefont{Anchordoqui et~al.}(2007)\citenamefont{Anchordoqui,
  Beacom, Goldberg, Palomares-Ruiz, and Weiler}}]{Anchordoqui:2006pe}
\bibinfo{author}{\bibfnamefont{L.~A.} \bibnamefont{Anchordoqui}},
  \bibinfo{author}{\bibfnamefont{J.~F.} \bibnamefont{Beacom}},
  \bibinfo{author}{\bibfnamefont{H.}~\bibnamefont{Goldberg}},
  \bibinfo{author}{\bibfnamefont{S.}~\bibnamefont{Palomares-Ruiz}},
  \bibnamefont{and} \bibinfo{author}{\bibfnamefont{T.~J.}
  \bibnamefont{Weiler}}, \bibinfo{journal}{Phys. Rev.}
  \textbf{\bibinfo{volume}{D75}}, \bibinfo{pages}{063001}
  (\bibinfo{year}{2007}), \eprint{astro-ph/0611581}.

\bibitem[{\citenamefont{Gonzalez-Garcia
  et~al.}(2012)\citenamefont{Gonzalez-Garcia, Maltoni, Salvado, and
  Schwetz}}]{GonzalezGarcia:2012sz}
\bibinfo{author}{\bibfnamefont{M.}~\bibnamefont{Gonzalez-Garcia}},
  \bibinfo{author}{\bibfnamefont{M.}~\bibnamefont{Maltoni}},
  \bibinfo{author}{\bibfnamefont{J.}~\bibnamefont{Salvado}}, \bibnamefont{and}
  \bibinfo{author}{\bibfnamefont{T.}~\bibnamefont{Schwetz}},
  \bibinfo{journal}{JHEP} \textbf{\bibinfo{volume}{1212}}, \bibinfo{pages}{123}
  (\bibinfo{year}{2012}), \eprint{1209.3023}.

\bibitem[{\citenamefont{Loeb and Waxman}(2006)}]{Loeb:2006tw}
\bibinfo{author}{\bibfnamefont{A.}~\bibnamefont{Loeb}} \bibnamefont{and}
  \bibinfo{author}{\bibfnamefont{E.}~\bibnamefont{Waxman}},
  \bibinfo{journal}{JCAP} \textbf{\bibinfo{volume}{0605}}, \bibinfo{pages}{003}
  (\bibinfo{year}{2006}), \eprint{astro-ph/0601695}.

\bibitem[{\citenamefont{Anchordoqui
  et~al.}(2014{\natexlab{b}})\citenamefont{Anchordoqui, Paul, da~Silva, Torres,
  and Vlcek}}]{Anchordoqui:2014yva}
\bibinfo{author}{\bibfnamefont{L.~A.} \bibnamefont{Anchordoqui}},
  \bibinfo{author}{\bibfnamefont{T.~C.} \bibnamefont{Paul}},
  \bibinfo{author}{\bibfnamefont{L.~H.~M.} \bibnamefont{da~Silva}},
  \bibinfo{author}{\bibfnamefont{D.~F.} \bibnamefont{Torres}},
  \bibnamefont{and} \bibinfo{author}{\bibfnamefont{B.~J.} \bibnamefont{Vlcek}},
  \bibinfo{journal}{Phys. Rev.} \textbf{\bibinfo{volume}{D89}},
  \bibinfo{pages}{127304} (\bibinfo{year}{2014}{\natexlab{b}}),
  \eprint{1405.7648}.

\bibitem[{\citenamefont{Tamborra et~al.}(2014)\citenamefont{Tamborra, Ando, and
  Murase}}]{Tamborra:2014xia}
\bibinfo{author}{\bibfnamefont{I.}~\bibnamefont{Tamborra}},
  \bibinfo{author}{\bibfnamefont{S.}~\bibnamefont{Ando}}, \bibnamefont{and}
  \bibinfo{author}{\bibfnamefont{K.}~\bibnamefont{Murase}}
  (\bibinfo{year}{2014}), \eprint{1404.1189}.

\bibitem[{\citenamefont{Chang and Wang}(2014)}]{Chang:2014hua}
\bibinfo{author}{\bibfnamefont{X.-C.} \bibnamefont{Chang}} \bibnamefont{and}
  \bibinfo{author}{\bibfnamefont{X.-Y.} \bibnamefont{Wang}}
  (\bibinfo{year}{2014}), \eprint{1406.1099}.

\bibitem[{\citenamefont{H{\"u}mmer et~al.}(2012)\citenamefont{H{\"u}mmer,
  Baerwald, and Winter}}]{Hummer:2011ms}
\bibinfo{author}{\bibfnamefont{S.}~\bibnamefont{H{\"u}mmer}},
  \bibinfo{author}{\bibfnamefont{P.}~\bibnamefont{Baerwald}}, \bibnamefont{and}
  \bibinfo{author}{\bibfnamefont{W.}~\bibnamefont{Winter}},
  \bibinfo{journal}{Phys. Rev. Lett.} \textbf{\bibinfo{volume}{108}},
  \bibinfo{pages}{231101} (\bibinfo{year}{2012}), \eprint{1112.1076}.

\bibitem[{\citenamefont{Klein et~al.}(2012)\citenamefont{Klein, Mikkelsen, and
  Tjus}}]{Klein:2012ug}
\bibinfo{author}{\bibfnamefont{S.~R.} \bibnamefont{Klein}},
  \bibinfo{author}{\bibfnamefont{R.}~\bibnamefont{Mikkelsen}},
  \bibnamefont{and} \bibinfo{author}{\bibfnamefont{J.~K.~B.}
  \bibnamefont{Tjus}} (\bibinfo{year}{2012}), \eprint{1208.2056}.

\bibitem[{\citenamefont{Winter et~al.}(2014)\citenamefont{Winter, Tjus, and
  Klein}}]{Winter:2014tta}
\bibinfo{author}{\bibfnamefont{W.}~\bibnamefont{Winter}},
  \bibinfo{author}{\bibfnamefont{J.~B.} \bibnamefont{Tjus}}, \bibnamefont{and}
  \bibinfo{author}{\bibfnamefont{S.~R.} \bibnamefont{Klein}}
  (\bibinfo{year}{2014}), \eprint{1403.0574}.

\bibitem[{\citenamefont{Reynoso}(2014)}]{Reynoso:2014yoa}
\bibinfo{author}{\bibfnamefont{M.~M.} \bibnamefont{Reynoso}}
  (\bibinfo{year}{2014}), \eprint{1403.3020}.

\bibitem[{\citenamefont{Hopkins and Beacom}(2006)}]{Hopkins:2006bw}
\bibinfo{author}{\bibfnamefont{A.~M.} \bibnamefont{Hopkins}} \bibnamefont{and}
  \bibinfo{author}{\bibfnamefont{J.~F.} \bibnamefont{Beacom}},
  \bibinfo{journal}{Astrophys. J.} \textbf{\bibinfo{volume}{651}},
  \bibinfo{pages}{142} (\bibinfo{year}{2006}), \eprint{astro-ph/0601463}.

\bibitem[{\citenamefont{Laha et~al.}(2013)\citenamefont{Laha, Beacom, Dasgupta,
  Horiuchi, and Murase}}]{Laha:2013lka}
\bibinfo{author}{\bibfnamefont{R.}~\bibnamefont{Laha}},
  \bibinfo{author}{\bibfnamefont{J.~F.} \bibnamefont{Beacom}},
  \bibinfo{author}{\bibfnamefont{B.}~\bibnamefont{Dasgupta}},
  \bibinfo{author}{\bibfnamefont{S.}~\bibnamefont{Horiuchi}}, \bibnamefont{and}
  \bibinfo{author}{\bibfnamefont{K.}~\bibnamefont{Murase}}
  (\bibinfo{year}{2013}), \eprint{1306.2309}.

\bibitem[{\citenamefont{Abbasi et~al.}(2011{\natexlab{a}})}]{Abbasi:2010ie}
\bibinfo{author}{\bibfnamefont{R.}~\bibnamefont{Abbasi}} \bibnamefont{et~al.}
  (\bibinfo{collaboration}{IceCube Collaboration}),
  \bibinfo{journal}{Phys. Rev.} \textbf{\bibinfo{volume}{D83}},
  \bibinfo{pages}{012001} (\bibinfo{year}{2011}{\natexlab{a}}),
  \eprint{1010.3980}.

\bibitem[{\citenamefont{Abbasi et~al.}(2011{\natexlab{b}})}]{Abbasi:2011jx}
\bibinfo{author}{\bibfnamefont{R.}~\bibnamefont{Abbasi}} \bibnamefont{et~al.}
  (\bibinfo{collaboration}{IceCube Collaboration}),
  \bibinfo{journal}{Phys. Rev.} \textbf{\bibinfo{volume}{D84}},
  \bibinfo{pages}{082001} (\bibinfo{year}{2011}{\natexlab{b}}),
  \eprint{1104.5187}.

\bibitem[{\citenamefont{Sinegovskaya et~al.}(2013)\citenamefont{Sinegovskaya,
  Ogorodnikova, and Sinegovsky}}]{Sinegovskaya:2013wgm}
\bibinfo{author}{\bibfnamefont{T.}~\bibnamefont{Sinegovskaya}},
  \bibinfo{author}{\bibfnamefont{E.}~\bibnamefont{Ogorodnikova}},
  \bibnamefont{and}
  \bibinfo{author}{\bibfnamefont{S.}~\bibnamefont{Sinegovsky}}
  (\bibinfo{year}{2013}), \eprint{1306.5907}.

\bibitem[{\citenamefont{Mena et~al.}(2014)\citenamefont{Mena, Palomares-Ruiz,
  and Vincent}}]{Mena:2014sja}
\bibinfo{author}{\bibfnamefont{O.}~\bibnamefont{Mena}},
  \bibinfo{author}{\bibfnamefont{S.}~\bibnamefont{Palomares-Ruiz}},
  \bibnamefont{and} \bibinfo{author}{\bibfnamefont{A.~C.}
  \bibnamefont{Vincent}} (\bibinfo{year}{2014}), \eprint{1404.0017}.

\bibitem[{\citenamefont{Chen et~al.}(2014)\citenamefont{Chen, Dev, and
  Soni}}]{Chen:2013dza}
\bibinfo{author}{\bibfnamefont{C.-Y.} \bibnamefont{Chen}},
  \bibinfo{author}{\bibfnamefont{P.~S.~B.} \bibnamefont{Dev}},
  \bibnamefont{and} \bibinfo{author}{\bibfnamefont{A.}~\bibnamefont{Soni}},
  \bibinfo{journal}{Phys. Rev.} \textbf{\bibinfo{volume}{D89}},
  \bibinfo{pages}{033012} (\bibinfo{year}{2014}), \eprint{1309.1764}.

\bibitem[{\citenamefont{Murase et~al.}(2013)\citenamefont{Murase, Ahlers, and
  Lacki}}]{Murase:2013rfa}
\bibinfo{author}{\bibfnamefont{K.}~\bibnamefont{Murase}},
  \bibinfo{author}{\bibfnamefont{M.}~\bibnamefont{Ahlers}}, \bibnamefont{and}
  \bibinfo{author}{\bibfnamefont{B.~C.} \bibnamefont{Lacki}},
  \bibinfo{journal}{Phys. Rev.} \textbf{\bibinfo{volume}{D88}},
  \bibinfo{pages}{121301} (\bibinfo{year}{2013}), \eprint{1306.3417}.

\bibitem[{\citenamefont{Abdo et~al.}(2010)}]{Abdo:2010nz}
\bibinfo{author}{\bibfnamefont{A.}~\bibnamefont{Abdo}} \bibnamefont{et~al.}
  (\bibinfo{collaboration}{Fermi-LAT collaboration}), \bibinfo{journal}{Phys.
  Rev. Lett.} \textbf{\bibinfo{volume}{104}}, \bibinfo{pages}{101101}
  (\bibinfo{year}{2010}), \bibinfo{note}{(see also update: talk by M. Ackermann
  at Fermi symposium, Monterey, USA, 2012)}, \eprint{1002.3603}.

\bibitem[{\citenamefont{Learned and Pakvasa}(1995)}]{Learned:1994wg}
\bibinfo{author}{\bibfnamefont{J.~G.} \bibnamefont{Learned}} \bibnamefont{and}
  \bibinfo{author}{\bibfnamefont{S.}~\bibnamefont{Pakvasa}},
  \bibinfo{journal}{Astropart. Phys.} \textbf{\bibinfo{volume}{3}},
  \bibinfo{pages}{267} (\bibinfo{year}{1995}), \eprint{hep-ph/9405296}.

\bibitem[{\citenamefont{Hillas}(1984)}]{Hillas:1985is}
\bibinfo{author}{\bibfnamefont{A.~M.} \bibnamefont{Hillas}},
  \bibinfo{journal}{Ann. Rev. Astron. Astrophys.}
  \textbf{\bibinfo{volume}{22}}, \bibinfo{pages}{425} (\bibinfo{year}{1984}).

\bibitem[{\citenamefont{Anchordoqui
  et~al.}(2014{\natexlab{c}})\citenamefont{Anchordoqui, Goldberg, Lynch,
  Olinto, Paul et~al.}}]{Anchordoqui:2013qsi}
\bibinfo{author}{\bibfnamefont{L.~A.} \bibnamefont{Anchordoqui}},
  \bibinfo{author}{\bibfnamefont{H.}~\bibnamefont{Goldberg}},
  \bibinfo{author}{\bibfnamefont{M.~H.} \bibnamefont{Lynch}},
  \bibinfo{author}{\bibfnamefont{A.~V.} \bibnamefont{Olinto}},
  \bibinfo{author}{\bibfnamefont{T.~C.} \bibnamefont{Paul}},
  \bibnamefont{et~al.}, \bibinfo{journal}{Phys. Rev.}
  \textbf{\bibinfo{volume}{D89}}, \bibinfo{pages}{083003}
  (\bibinfo{year}{2014}{\natexlab{c}}), \eprint{1306.5021}.

\bibitem[{\citenamefont{Murase et~al.}(2008)\citenamefont{Murase, Ioka,
  Nagataki, and Nakamura}}]{Murase:2008mr}
\bibinfo{author}{\bibfnamefont{K.}~\bibnamefont{Murase}},
  \bibinfo{author}{\bibfnamefont{K.}~\bibnamefont{Ioka}},
  \bibinfo{author}{\bibfnamefont{S.}~\bibnamefont{Nagataki}}, \bibnamefont{and}
  \bibinfo{author}{\bibfnamefont{T.}~\bibnamefont{Nakamura}},
  \bibinfo{journal}{Phys. Rev.} \textbf{\bibinfo{volume}{D78}},
  \bibinfo{pages}{023005} (\bibinfo{year}{2008}), \eprint{0801.2861}.

\bibitem[{\citenamefont{Razzaque et~al.}(2004)\citenamefont{Razzaque, Meszaros,
  and Waxman}}]{Razzaque:2004yv}
\bibinfo{author}{\bibfnamefont{S.}~\bibnamefont{Razzaque}},
  \bibinfo{author}{\bibfnamefont{P.}~\bibnamefont{Meszaros}}, \bibnamefont{and}
  \bibinfo{author}{\bibfnamefont{E.}~\bibnamefont{Waxman}},
  \bibinfo{journal}{Phys. Rev. Lett.} \textbf{\bibinfo{volume}{93}},
  \bibinfo{pages}{181101} (\bibinfo{year}{2004}), \eprint{astro-ph/0407064}.

\bibitem[{\citenamefont{Ando and Beacom}(2005)}]{Ando:2005xi}
\bibinfo{author}{\bibfnamefont{S.}~\bibnamefont{Ando}} \bibnamefont{and}
  \bibinfo{author}{\bibfnamefont{J.~F.} \bibnamefont{Beacom}},
  \bibinfo{journal}{Phys. Rev. Lett.} \textbf{\bibinfo{volume}{95}},
  \bibinfo{pages}{061103} (\bibinfo{year}{2005}), \eprint{astro-ph/0502521}.

\bibitem[{\citenamefont{Razzaque et~al.}(2005)\citenamefont{Razzaque, Meszaros,
  and Waxman}}]{Razzaque:2005bh}
\bibinfo{author}{\bibfnamefont{S.}~\bibnamefont{Razzaque}},
  \bibinfo{author}{\bibfnamefont{P.}~\bibnamefont{Meszaros}}, \bibnamefont{and}
  \bibinfo{author}{\bibfnamefont{E.}~\bibnamefont{Waxman}},
  \bibinfo{journal}{Mod. Phys. Lett.} \textbf{\bibinfo{volume}{A20}},
  \bibinfo{pages}{2351} (\bibinfo{year}{2005}), \eprint{astro-ph/0509729}.

\bibitem[{\citenamefont{Tjus et~al.}(2014)\citenamefont{Tjus, Eichmann, Halzen,
  Kheirandish, and Saba}}]{Tjus:2014dna}
\bibinfo{author}{\bibfnamefont{J.~B.} \bibnamefont{Tjus}},
  \bibinfo{author}{\bibfnamefont{B.}~\bibnamefont{Eichmann}},
  \bibinfo{author}{\bibfnamefont{F.}~\bibnamefont{Halzen}},
  \bibinfo{author}{\bibfnamefont{A.}~\bibnamefont{Kheirandish}},
  \bibnamefont{and} \bibinfo{author}{\bibfnamefont{S.}~\bibnamefont{Saba}}
  (\bibinfo{year}{2014}), \eprint{1406.0506}.

\bibitem[{\citenamefont{Lemoine et~al.}(2006)\citenamefont{Lemoine, Pelletier,
  and Revenu}}]{Lemoine:2006gg}
\bibinfo{author}{\bibfnamefont{M.}~\bibnamefont{Lemoine}},
  \bibinfo{author}{\bibfnamefont{G.}~\bibnamefont{Pelletier}},
  \bibnamefont{and} \bibinfo{author}{\bibfnamefont{B.}~\bibnamefont{Revenu}},
  \bibinfo{journal}{Astrophys. J.} \textbf{\bibinfo{volume}{645}},
  \bibinfo{pages}{L129} (\bibinfo{year}{2006}), \eprint{astro-ph/0606005}.

\bibitem[{\citenamefont{Kachelriess and Semikoz}(2006)}]{Kachelriess:2005xh}
\bibinfo{author}{\bibfnamefont{M.}~\bibnamefont{Kachelriess}} \bibnamefont{and}
  \bibinfo{author}{\bibfnamefont{D.~V.} \bibnamefont{Semikoz}},
  \bibinfo{journal}{Phys. Lett.} \textbf{\bibinfo{volume}{B634}},
  \bibinfo{pages}{143} (\bibinfo{year}{2006}), \eprint{astro-ph/0510188}.

\bibitem[{\citenamefont{Ahlers et~al.}(2011)\citenamefont{Ahlers,
  Gonzalez-Garcia, and Halzen}}]{Ahlers:2011jj}
\bibinfo{author}{\bibfnamefont{M.}~\bibnamefont{Ahlers}},
  \bibinfo{author}{\bibfnamefont{M.}~\bibnamefont{Gonzalez-Garcia}},
  \bibnamefont{and} \bibinfo{author}{\bibfnamefont{F.}~\bibnamefont{Halzen}},
  \bibinfo{journal}{Astropart. Phys.} \textbf{\bibinfo{volume}{35}},
  \bibinfo{pages}{87} (\bibinfo{year}{2011}), \eprint{1103.3421}.

\bibitem[{\citenamefont{Baerwald et~al.}(2015)\citenamefont{Baerwald,
  Bustamante, and Winter}}]{Baerwald:2014zga}
\bibinfo{author}{\bibfnamefont{P.}~\bibnamefont{Baerwald}},
  \bibinfo{author}{\bibfnamefont{M.}~\bibnamefont{Bustamante}},
  \bibnamefont{and} \bibinfo{author}{\bibfnamefont{W.}~\bibnamefont{Winter}},
  \bibinfo{journal}{Astropart. Phys.} \textbf{\bibinfo{volume}{62}},
  \bibinfo{pages}{66} (\bibinfo{year}{2015}), \eprint{1401.1820}.

\bibitem[{\citenamefont{Katz et~al.}(2013)\citenamefont{Katz, Waxman, Thompson,
  and Loeb}}]{Katz:2013ooa}
\bibinfo{author}{\bibfnamefont{B.}~\bibnamefont{Katz}},
  \bibinfo{author}{\bibfnamefont{E.}~\bibnamefont{Waxman}},
  \bibinfo{author}{\bibfnamefont{T.}~\bibnamefont{Thompson}}, \bibnamefont{and}
  \bibinfo{author}{\bibfnamefont{A.}~\bibnamefont{Loeb}}
  (\bibinfo{year}{2013}), \eprint{1311.0287}.

\bibitem[{\citenamefont{Abbasi et~al.}(2010)}]{Abbasi:2009nf}
\bibinfo{author}{\bibfnamefont{R.}~\bibnamefont{Abbasi}} \bibnamefont{et~al.}
  (\bibinfo{collaboration}{HiRes Collaboration}),
  \bibinfo{journal}{Phys. Rev. Lett.} \textbf{\bibinfo{volume}{104}},
  \bibinfo{pages}{161101} (\bibinfo{year}{2010}), \eprint{0910.4184}.

\bibitem[{\citenamefont{Liu et~al.}(2014)\citenamefont{Liu, Wang, Inoue,
  Crocker, and Aharonian}}]{Liu:2013wia}
\bibinfo{author}{\bibfnamefont{R.-Y.} \bibnamefont{Liu}},
  \bibinfo{author}{\bibfnamefont{X.-Y.} \bibnamefont{Wang}},
  \bibinfo{author}{\bibfnamefont{S.}~\bibnamefont{Inoue}},
  \bibinfo{author}{\bibfnamefont{R.}~\bibnamefont{Crocker}}, \bibnamefont{and}
  \bibinfo{author}{\bibfnamefont{F.}~\bibnamefont{Aharonian}},
  \bibinfo{journal}{Phys. Rev.} \textbf{\bibinfo{volume}{D89}},
  \bibinfo{pages}{083004} (\bibinfo{year}{2014}), \eprint{1310.1263}.

\bibitem[{\citenamefont{Kistler et~al.}(2013)\citenamefont{Kistler, Stanev, and
  Yuksel}}]{Kistler:2013my}
\bibinfo{author}{\bibfnamefont{M.~D.} \bibnamefont{Kistler}},
  \bibinfo{author}{\bibfnamefont{T.}~\bibnamefont{Stanev}}, \bibnamefont{and}
  \bibinfo{author}{\bibfnamefont{H.}~\bibnamefont{Yuksel}}
  (\bibinfo{year}{2013}), \eprint{1301.1703v2}.

\end{thebibliography}

\end{document}